\documentclass[aip,apl,reprint,preprintnumbers,
]{revtex4-1}
\usepackage{graphicx}
\usepackage{dcolumn}
\usepackage{bm}
\usepackage{color}
\usepackage{sidecap}
\usepackage{amssymb}

\begin{document}

\title{Reply to ``Comment on `Resilience of gated avalanche photodiodes against bright illumination attacks in quantum cryptography'"}
\author {Z. L. Yuan}
\email{zhiliang.yuan@crl.toshiba.co.uk}
\author {J. F. Dynes}
\author {A. J. Shields}
\address{Toshiba Research Europe Ltd, Cambridge Research Laboratory, 208 Cambridge Science Park, Milton Road, Cambridge, CB4~0GZ, UK }

\date{\today}

\maketitle

Quantum key distribution (QKD) has been proven theoretically secure at the protocol level. However, the security may be compromised through deviation from theoretical models in device implementation or operation.  For each deviation, thorough understanding must be achieved in order for subsequent construction of robust countermeasures.

In Ref.~\onlinecite{yuan11}, we have studied the effectiveness of bright illumination attacks, a group of attacks targeting gated InGaAs avalanche photodiodes (APDs).\cite{lydersen10,lydersen10b} We found that gated APDs are naturally resilient against continuous-wave (CW) attacks through the gain modulation effect.\cite{yuan10b} The finding is contrary to the claim by Lydersen \textit{et al.}\cite{lydersen10} that ``the loophole is likely to be present in most QKD systems using APDs to detect single photons". The detector loophole reported by Lydersen \textit{et al.}\cite{lydersen10} was in fact due to inappropriate settings in the discrimination level of single photon APDs.  Furthermore, we discussed the respective effectiveness for temporally-tailored bright illumination\cite{lydersen10b} and after-gate attacks,\cite{wiechers11} as summarized in Table I. Against bright illumination attacks, ``monitoring the photocurrent" was proposed as a counter-measure, which was based on the measured difference in the APD currents as compared with normal operation, see Fig.~1.

In the preceding comment,\cite{lydersen11} we note that Lydersen \textit{et al.} do not dispute that bright light attacks are ineffective if the detector parameters are set correctly, which was our main finding. Instead, they challenge the robustness of the counter-measure we proposed. To serve this challenge, they have designed a new attack\cite{lydersen11b} that uses faint optical pulses ($\le$120 photons/pulse) only. The attack exploits the super-linear count dependence achieved by restricting the avalanche duration.\cite{kardynal08,yuan10c} In the absence of bright illumination, Lydersen \textit{et al.} claim, unfortunately without experimental elaboration, that the attack ``would not be detectable"\cite{lydersen11} by monitoring the current and ``the afterpulsing is negligible".\cite{lydersen11b}
It should be pointed out that Lydersen \textit{et al.} reported a high quantum bit error ratio (QBER) of $>$12\% during the attack, which would be easily detected by the legitimate users of the system, even for a 100 kHz clock rate which significantly favours the attacker.

\begin{figure}[b]
\newpage
\centering
\includegraphics[width=.96\columnwidth]{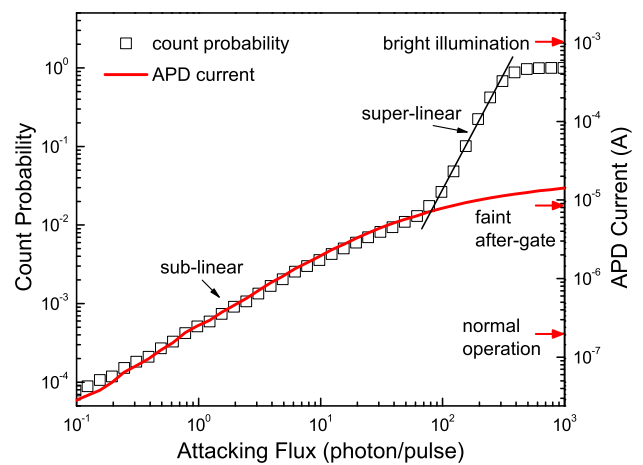}
\caption{Count probability and photocurrent when under the faint after-gate attack as a function of photon flux. Arrows in the right axis label respective currents measured for normal operation, faint after-gate attack and bright illumination attacks.}
\label{fig:faintattck}
\newpage
\end{figure}

Although the illumination during the attack is weak, Lydersen \textit{et al.} have overlooked the fact that the APD gain is still large, and thus the attack generates a sizable photocurrent that can easily be detected by the users.  We prove this with a simple experiment. A gated InGaAs APD is subjected to faint illumination by a 50 ps pulsed laser. The APD is gated with 2~MHz pulses of 4~V amplitude and 3.5~ns duration, a gating condition identical to previously used in an actual QKD system.\cite{gobby04} Its single photon detection efficiency is measured to be 15\% with a dark count probability of $3.5\times10^{-5}$ per gate. In the attack, the pulse delay is optimized to have a minimum QBER that the attack would have caused with a flux of 120 photons. Under this optimized delay, the count probability and photocurrent are measured as a function of attacking flux, as shown in Fig.~1. Super-linearity regime used in the faint after-gate attack is identified as occurring at fluxes between 70-300 photons/pulse. Within this flux range, the APD current is macroscopic, and actually easily detectable. At around 9~$\mu$A, this current is more than 40 times stronger than would be under normal QKD operation for this APD,\cite{normal} see Fig.~1.

Macroscopic current causes afterpulsing.\cite{yuan10} In Fig.~1, for an attacking flux less than 60 photons/pulse, the count probability shows a slightly sub-linear dependence, closely resembling that of the photocurrent. This sub-linear behavior can be explained only by the dominance of afterpulsing, and this assignment has been verified by the gated afterpulse measurement technique.\cite{bethune04,yuan07,namekata09}  As can be extrapolated from the linear dependence, the afterpulsing is \textit{not} negligible in the super-linear regime.
Therefore, contrary to the intuitive claims by Lydersen \textit{et al.}, the faint after-gate attack not only causes significant current but also produces non-negligible afterpulses. In addition to the resultant QBER,\cite{lydersen11b} monitoring the APD current is an effective counter-measure against this attack, although it was initially proposed for bright illumination attacks.\cite{yuan11}

The experiment shown in Fig.~1 illustrates again the importance of careful analysis of any proposed attack. Careful analysis is the very foundation, upon which a robust, effective counter-measure can be constructed. Against an attack, two strategies are usually adopted: (i) bounding the information leakage, followed by privacy amplification; and (ii) deterministic detection or exclusion. Only attacks that are not easily detected on a properly implemented system need to be accounted for in the privacy amplification analysis, \textit{e.g.}, the photon number splitting attack.\cite{wang05,lo05}  This is certainly not the case for the various detector blinding attacks which rely upon a poor design of the APD circuit and which generate a large, easily-detectable photocurrent in the device.

\begin{table*}[t]
\caption{Summary of attacks targeting the photocurrent mode of gated InGaAs APDs.}
\begin{ruledtabular}
\begin{tabular}{p{3cm}|p{4.7cm}|p{4.7cm}|p{4.7cm}}
  \textbf{Attack}& \textbf{Effectiveness} (without counter-measure) & \textbf{Fingerprint} & \textbf{Counter-measure} \\
\hline CW blinding\cite{lydersen10} & Ineffective to correctly operated devices & High photocurrent & Monitor the current\\
\hline CW thermal blinding\cite{lydersen10b} & Ineffective to correctly operated devices  & High photocurrent & Monitor the current\\
\hline Thermal blinding of frames\cite{lydersen10b}  & Limited effectiveness to burst-mode systems & High photocurrent; Giveaway photon clicks & Monitor the current\\
\hline sinkhole blinding\cite{lydersen10b} & Limited effectiveness to APDs with an AC-coupled output; Ineffective to DC-coupled APDs & High photocurrent; Giveaway photon clicks & Monitor the current\\
\hline After-gate\cite{wiechers11} & Limited effectiveness to burst-mode systems & Photon arrival timing; High QBER due to afterpulses & Use of narrow modulation and/or detection acceptance window\\
\hline\hline Faint after-gate\cite{lydersen11b} & Ineffective due to high QBER & High photocurrent; High QBER due to finite count super-linearity and afterpulses & Monitor the current
\end{tabular}
\end{ruledtabular}
\label{tab:attacks}
\end{table*}

\newpage

%


\begin{thebibliography}{17}
\bibitem [{nor()}]{normal}%
  \BibitemOpen
  \href@noop {} {}\bibinfo {note} {The APD current is measured to be 200~nA for
  an arrival flux of 0.1 photons/pulse at 2~MHz in the single-photon counting
  setting. The arrival flux is appropriate as the upper limit under normal
  operation.}\BibitemShut {Stop}%
\end{thebibliography}

\end{document}